\documentclass[12pt]{article}
\usepackage{setspace}
\usepackage[normalem]{ulem}
\usepackage{geometry}
\usepackage{lineno}
\usepackage{graphicx}
\usepackage{amssymb}
\usepackage{amsmath}
\usepackage{indentfirst}
\usepackage{natbib}
\usepackage{color}
\definecolor{darkgreen}{cmyk}{0.85,0.2,1.00,0.2}

\def\barray{\begin{array}}
\def\earray{\end{array}}
\def\be{\begin{equation}}
\def\ee{\end{equation}}
\def\ben{\begin{equation} \nonumber}
\def\een{\end{equation}}
\def\ban{\begin{eqnarray*}}
\def\ean{\end{eqnarray*}}
\def\ba{\begin{eqnarray}}
\def\ea{\end{eqnarray}}
\def\pcite[#1]{(\citealt{#1})}
\def\egcite[#1]{(e.g., \citealt{#1})}
\def\egciteParens[#1]{(e.g., \citealt{#1})}

\def\({\left(}
\def\){\right)}
\def\half{{1\over2}}

\begin{document}
\thispagestyle{empty}
\begin{tabbing}
TITLE: \hspace{1 in} \= Sex-specific recombination rates and allele frequencies  \\
\> affect the invasion of sexually antagonistic variation on autosomes\\
\\
ARTICLE TYPE: \> Research article\\
\\
RUNNING HEAD: \> Sex-specific recombination rates and allele frequencies\\
\\
AUTHORS: \> {Minyoung J. Wyman$^{1,2}$} \\
\> {Mark C. Wyman$^{3,4}$} \\
\\
ADDRESS:
\>$^{1}$minyoung.yi@mail.utoronto.ca \\
\>$^{2}$Department of Ecology \& Evolutionary Biology\\
\>University of Toronto \\
\>25 Willcocks Street \\
\>Toronto, ON M5S 3B2 \\
\>CANADA\\
\\
\>$^{3}$Department of Astronomy \& Astrophysics\\
\>$^4$Kavli Institute for Cosmological Physics, Enrico Fermi Institute\\
\>University of Chicago\\
\>Chicago, IL 60637\\
\>USA
\end{tabbing}

\pagebreak
\section{Abstract}
\raggedright
\parindent=3.5em 
The introduction and persistence of novel sexually antagonistic alleles can depend upon factors that differ between males and females.   Understanding the conditions for invasion in a two-locus model can elucidate these processes.  For instance,  selection can act differently upon the sexes, or sex-linkage can facilitate the invasion of genetic variation with opposing fitness effects between the sexes.  Two factors that deserve further attention are recombination rates and allele frequencies -- both of which can vary substantially between the sexes.  We find that sex-specific recombination rates in a two-locus diploid model can affect the invasion outcome of sexually antagonistic alleles and that the sex-averaged recombination rate is not necessarily sufficient to predict invasion. We confirm that the range of permissible recombination rates is smaller in the sex benefitting from invasion and larger in the sex harmed by invasion. However, within the invasion space, male recombination rate can be greater than, equal to, or less than female recombination rate in order for a male-benefit, female-detriment allele to invade (and similarly for a female-benefit, male-detriment allele).  We further show that a novel, sexually antagonistic allele that is also associated with a lowered recombination rate can invade more easily when present in the double heterozygote genotype. Finally, we find that sexual dimorphism in resident allele frequencies can impact the invasion of new sexually antagonistic alleles at a second locus.   Our results suggest that accounting for sex-specific recombination rates and allele frequencies can determine the difference between invasion and non-invasion of novel sexually antagonistic alleles in a two-locus model.

\pagebreak
\setcounter{page}{1}
\section{Introduction}
\raggedright
\parindent=3.5em 
Selection can act differently in males versus females \pcite[arnqvistRowe2005], resulting in antagonism over the expression of shared traits.  Fitness itself can be under sexual conflict so that reproductive success has a negative genetic correlation between the sexes \egcite[chippindaleEtAl2001, foersterEtAl2007].  Alternatively, individual traits may also be under sexual conflict because trait expression in one sex has opposing fitness effects in the other sex.  Among sexually dimorphic traits, $\sim$17\% of selection estimates were sexually antagonistic \pcite[coxCalsbeek2009]. Sexual conflict can operate at even finer scale levels: the expression of $\sim$8\% of genes is beneficial for one sex but detrimental in the other in \emph{Drosophila melanogaster} \pcite[innocentiMorrow]. Understanding sexual antagonism requires experimental \egcite[longEtAl2012] as well as theoretical approaches. In a one-locus model, \cite{rice84} suggested that sex-linkage can facilitate the initial spread of sexually antagonistic alleles. His model demonstrated that recessivity and hemizygosity (possessing one major sex chromosome) can shield novel antagonistic alleles from selection in the non-benefitting sex. However, \cite{fry} has shown that differences in sex-specific dominance may affect the invasion advantage of sex-linkage.

Recent theoretical work on the invasion of sexually antagonistic alleles has focused on two-locus models \pcite[cc2010, pattenEtAl2010, ubedaEtAl2011] which introduce realistic factors such as epistasis, linkage disequilibrium, and recombination. These factors are known to alter the invasion outcomes for novel sexually antagonistic alleles. For instance, \cite{cc2010} employed a model whereby sexual antagonism can be introduced through epistatic interactions between two different loci. When epistasis is present, the recombination rate difference between males and females affects the invasion of sexually antagonistic allelic combinations. In a related model, \cite{pattenEtAl2010} showed that a second locus can both increase the opportunities for polymorphic equilibria relative to a one-locus model and preserve allelic combinations that increase fitness variation through linkage disequilibrium.

Differences in sex-specific recombination rates are often dramatic \egcite[mankRecomb, lenorDuth, brandvainCoop2012], suggesting the need for further study into its implications. Up to 75\% of recombining species demonstrate $>$5\% overall rate difference between the sexes \pcite[burt, lenormand]. Recombination rates on a local scale can also vary among individuals \pcite[coopEtAl2008, baudatEtAl2010, fledelAlonEtAl2011] and between the sexes \egcite[hansson, perrin, kong]. In fact, recombination rates have very little intersexual covariance \pcite[coopEtAl2008], so that tracking the rates in males and females separately may be necessary.  It is important to understand what, if any, implications  sexual dimorphism in recombination might have on the introduction of sexually antagonistic phenotypes. For instance, it has been suggested that lower recombination can maintain combinations of genes beneficial to males because the genes have successfully undergone sexual selection \pcite[trivers1988].  Although sex-specific selection at the diploid stage is unlikely to facilitate the evolution of heterochiasmy \pcite[lenormand], once heterochiasmy is established, it may subsequently affect the formation of male- or female-benefit regions in the genome.  Indeed, lower sex-specific recombination makes it easier for alleles benefitting that same sex to invade on sex chromosomes \pcite[cc2010].

Here we expand upon previous models and investigate the effect of heterochiasmy in three ways. First, we study the effect of sexually dimorphic recombination rates on the invasion of sexually antagonistic alleles. Specifically, we examine whether dimorphism in recombination can hinder or facilitate the spread of an autosomal male-benefit, female-detriment allele (and similarly for female-benefit, male-detriment alleles). Second, to capture the reality of sex-specific and individual variation in recombination rates, we study how sex- and genotype-specific recombination rates affect invasion outcomes. Finally, we analyze the impact of sexually dimorphic allele frequencies at one locus for invasion outcome at a second locus. Considered together, these results suggest that novel sexually antagonistic variation can potentially spread more easily due to pre-existing sexual dimorphism in recombination and allele frequencies.

\section{Model}
We constructed an autosomal two-locus diploid model. The two-locus diploid model has been studied extensively in other contexts \egcite[otto, curt, albert, pattenEtAl2010]. Much of our notation is adopted from Connallon and Clark (2010) to facilitate comparison. The $A$ and $B$ loci both have two alleles so that there are 10 unique male and 10 unique female genotypes and associated fitnesses (see Table \ref{tab1}). Recombination between the $A$ and $B$ locus occurs at a rate $r_m$ in males and $r_f$ in females. Discrete time recursion equations for the $AB$ genotypes are reproduced in the Appendix \ref{recursions} (and in a {\tt Mathematica} notebook provided in the online supplementary materials). The genotype frequencies in the next generation depend upon the maternal and paternal allele frequencies at the $A$ and $B$ loci, as well as upon recombination and linkage disquilibrium between the two loci. 

First, we analyze recombination rates that are non-zero (i.e., no achiasmy) and different in each sex for the polymorphic equilibrium case (e.g., $A_1$ is not fixed upon the invasion of $B_2$). We first work with a general model to derive our expressions,  and then plug in a specific fitness parameterization to investigate the effects of dominance and epistasis on our results (parameterization in Appendix A). Because we only specialize at the last step, any fitness parameterization may be used, and our general results are not contingent upon them. Although we are using a two-locus diploid model in the context of sexual antagonism, our model is not to be conflated with models of interlocus sexual conflict (i.e., different loci in each sex interact to affect the expression of a shared trait). Second, we treat recombination rate as a novel sexually antagonistic trait in our model and study its effect on invasion. Lastly, we study the model's behavior on the invasion of new alleles when resident allele frequencies are different between the sexes. 

\section{Results and Discussion}
\subsection {Sex-averaged versus sex-specific recombination rates} \label{sexave}
Upon first consideration, it may appear that only the sex-averaged recombination rate can matter for invasion \egcite[hedrick]. After all, recombination shuffles loci in males and females, so that any particular allelic combination has approximately the same likelihood of appearing in both sons and daughters. These combinations experience sex-specific selection, but then get re-shuffled in males and females of the second generation. As re-shuffling is inevitable, it may seem that the effects of sex-specific recombination are averaged out between the sexes over time. However, a few lines of evidence suggest that this view is too simplistic.  First, sexual antagonism can preserve stable linkage disequilibrium in the face of recombination so that sex-specific patterns of variation persist as polymorphic equilibria \pcite[pattenEtAl2010, ubedaEtAl2011]. Second, when sex-specific recombination rates are explicitly modeled, their particular values appear to affect the conditions for invasion \pcite[cc2010]. Finally, whether a novel allele spreads or not depends only upon the present conditions, not the long-term conditions. Selection precedes fertilization, so that recombination occurs in a subset of individuals different from the total set initially preset at birth. Selection may occur at the adult stage, but may also occur at the gamete stage via fertility selection. Thus, the interplay of sexually antagonistic fitness and sex-specific recombination rates may lead to small but potentially consequential sex differences in the frequency of novel alleles, affecting their invasion.

We can study these dynamics by conducting a local stability analysis of the recursion equations governing a two-locus model with sex-specific recombination (Appendix \ref{recursions}). We calculated the eigenvalues of the Jacobian matrix for the recursion equations; whenever the dominant eigenvalue describing an invasive allele is greater than one, the population is subject to invasion by that novel allele. We conducted this analysis under the condition that the first $A$ locus is polymorphic for $A_1$ (frequency $q$) and $A_2$ (frequency $p$) and the second $B$ locus is fixed for $B_1$ upon the introduction of $B_2$. The polymorphic case may be more interesting under the assumption that a novel mutation may appear on a background that is genetically variable throughout a population (NB: The eigenvalue for the equilibrium whereby $A_1$ is fixed may be recovered by setting $p=0$ and $q=1$). 

 The dominant eigenvalue ($\lambda_{CC}$) for the invasion of a novel $B_2$ allele when the male and female frequencies are equal  ($p_m = p_f$) at the $A$ locus is given by equation 6a in Connallon and Clark (2010):
\ba
\lambda_{CC} = \frac{1}{2\,\bar{w}_f} \( f_{32} p + f_{22C} (1-p) (1-r_f) \) + \frac{1}{2\,\bar{w}_m} \( m_{32} p + m_{22C} (1-p) (1-r_m) \). \label{cceig}
\ea
As Connallon and Clark (2010), we have used $f_{\rm type}$ and $m_{\rm type}$ to designate female and male fitnesses, respectively (Table \ref{tab1}). For instance,  genotype $A_2B_1 A_2B_2$ has the fitness type 32. In addition, the 22 types are either 22C or 22R, because 22 is heterozygous at both loci, but may occur as coupling or repulsion (Table \ref{tab1}). $\bar{w}_f$ and $\bar w_m$ represent the mean female and male fitnesses, respectively. 

$\lambda_{CC}$ can be re-written by expressing fitness values as a deviation from 1. For instance, for the $A_2B_1 A_2B_2$ genotype  the female fitness can be re-expressed as $
f_{32} = 1 - u \; \delta f_{32}$, where
$\delta f_{32}$ is the deviation of $f_{32}$ from 1. The $u$ term is a counting parameter for doing a series expansion when $u$ is small.  
Terms with no $u$ (i.e., $u^0$) are larger than terms with $u^1$, and terms with $u^1$ are larger than terms with $u^2$, etc.  This expansion is made assuming weak selection; we also assume that mutation is weak. Finally, we further assume that recombination rates are small and also multiply them by $u$. Equation \ref{cceig} can now be re-written as an expansion in $u$:
\ba
\lambda_{CC} & \simeq & \half \( \frac{\bar{w}_f + \bar{w}_m}{\bar{w}_f \, \bar{w}_m} \) +  \nonumber \\
&& - \frac{u}{2 \bar{w}_f \bar{w}_m} \(\bar{w}_f \( (1-p) \delta m_{22C} + p \delta m_{32} \) + \bar{w}_m\((1-p) \delta f_{22C} + p \delta f_{32} \) + (1-p)(\bar{w}_m r_f + \bar{w}_f r_m) \) 
\nonumber \\
&& +  \frac{u^2}{2 \bar{w}_f \bar{w}_m} \(1-p\)  \( \delta f_{22C} \bar{w}_m r_f + \delta m_{22C} \bar{w}_f r_m \) + \mathcal{O}(u^3).
\label{cceigex1}
\ea
This  approximation has a few nice features that aid our understanding. The first line contains just the mean fitnesses and is approximately $1$.  The second line contains terms that have one power of $u$; the third line contains even smaller pieces (where $\mathcal{O}$ indicates the order in u). Taking $\bar{w}_f = \bar{w}_m = 1$ simplifies things further: 
\ba
\lambda_{CC} & \simeq & 1  - \frac{u}{2} [ (1-p) \delta m_{22C} + p \delta m_{32}  + (1-p) \delta f_{22C} + p \, \delta f_{32} +  (1-p)( r_f + r_m) ] 
\nonumber \\
&& +  \frac{u^2}{2} \(1-p\)  [ \delta f_{22C} r_f + \delta m_{22C} r_m ] + \mathcal{O}(u^3).
\label{ccsimp}
\ea

We observe that the main recombination effect is indeed the sex-averaged recombination rate (the last term in the first line of equation \ref{ccsimp}), as is typically assumed \pcite[hedrick]. However, 
when males and females differ in fitness for the genotype $A_1B_1A_2B_2$ -- i.e., when $\delta f_{22} \neq \delta m_{22}$ (as is expected when $B_2$ is sexually antagonistic) -- the sex-averaged rate is adjusted by the second line (i.e., $\mathcal{O}(u^2)$). The strength of the second line is proportional to $1-p$ and affects genotypes bearing $B_2$. As the strength of the sexual antagonism increases, the effect of heterochiasmy also increases.

In general, the $\mathcal{O}(u^2)$ terms will be smaller than the $\mathcal{O}(u^1)$ terms. However this assumption is not met in two cases. Most obviously, the approximation is not valid under strong selection or strong sexual antagonism (i.e., genotype fitnesses are not close to 1) so that the approximation $\mathcal{O}(u^2) \ll \mathcal{O}(u^1)$ breaks down.  A more subtle scenario (that does not violate the approximation) is when the terms on the first line almost cancel out, allowing order 1 terms to be order 2:
\ba
\frac{u}{2} [ (1-p) \delta m_{22C} + p \delta m_{32}  + (1-p) \delta f_{22C} + p \delta f_{32}  +  (1-p)( r_f + r_m)  ] \simeq u^2. \label{specialCase}
\ea
For example, for the stable polymorphic equilibrium $p=0.25$ [i.e., polymorphic equilibrium is determined by $p = (W_{AA} - W_{Aa}$) / ($2W_{Aa}-W_{AA}-W_{aa})$, in the notation of \cite{ottoDay}], the $\delta m$ terms are all slightly negative (male-benefit) and the $\delta f$ terms are all slightly positive (female-detriment). As a result, the magnitude of $ \delta m_{22} \simeq$ the magnitude of $ \delta f_{22}$, and the magnitude of $\delta m_{32}$ $\simeq$ magnitude of $\delta f_{32}$, but with the overall sum of the four terms having a slightly negative effect. In other words, the $A_2$ and $B_2$ alleles are only slightly male-benefit and slightly female-detriment. Such alleles may be important to study under the assumption that new mutations may have minor phenotypic effects in relation to the current phenotype, as we assume under an infinitesimal model.  Then the sum within the brackets of equation \ref{specialCase} may be very small, of $\mathcal{O}(u^1)$.  Since $r_m$ and $r_f$ are of  $\mathcal{O}(u^1)$, the ``first" order term is the same or smaller magnitude as the second order term.  In such a case, the second order term makes a substantial difference to the outcome of invasion and must be included for consistency; so, the sex-averaged rates and the relative rates may contribute equally when equation \ref{specialCase} holds. Even though this special case may only persist for a short time in a population at large, it can affect the invasion of new alleles during that time. (NB: We emphasize that when ${\cal O}(u) \simeq {\cal O}(u^2)$ due to an accidental cancellation, the approximation is still valid because we still expect ${\cal O}(u^3) \ll {\cal O}(u^2)$). We have demonstrated qualitatively how equation \ref{specialCase} may be relevant. Below we also present an example using a particular fitness model to highlight both the qualitative and quantitative implications of equation \ref{specialCase} to invasion.

\subsection{Implications of sex-specific recombination}
The analysis above shows that the sex-specific recombination rates matter for invasion, in addition to the sex-averaged rate. A specific fitness model (see Appendix A) can be used for further investigation. We determine whether $B_2$ can invade on a polymorphic $A$ background, given that $B_2$ decreases fitness in both males and females, but also interacts epistatically with $A_2$ to increase only male fitness and not alter female fitness (i.e., the $A_2B_2$ combination is sexually antagonistic overall). The fitness model 
assumes linear dominance (dominance coefficients $h$ and $g$ for the $A$ and $B$ loci, respectively). The selection coefficients  are $s$ and $t$ (for A and B loci, respectively); positive values indicate a fitness detriment and negative values indicate a fitness benefit. Epistasis ($\epsilon$) is additive; positive values indicate a fitness detriment and negative values indicate a fitness benefit.

 For the $A$ locus we assumed equal dominance and selection coefficients between the sexes: $h_f=-0.1$, $h_m=-0.1$, $s_f=0.1$, and $s_m=0.1$. For the $B$ locus we also assumed equal dominance and selection coefficients: $g_f=0.5$, $g_m=0.5$, $t_f=0.05$, and $t_m=0.05$. For these set of parameters, the $B_2$ allele is detrimental to both males and females. However, we assumed epistasis to be $\epsilon_m=-0.25$ and $\epsilon_f=0$. As a result, when $A_2$ and $B_2$ are together, the allele combination confers an overall benefit to males but not females (since $t_m=t_f$ but $\epsilon_m\neq\epsilon_f$). It would be easy enough to make the alleles themselves sexually antagonistic by setting selection to have opposite signs between the sexes (i.e., $s_m$, $t_m >$0 and $s_f$, $t_f<$0).  However, doing so at the $A$ locus would violate the assumption of equal allele frequencies in males and females (see more below). Meanwhile, making the $B$ locus sexually antagonistic independently of the $A$ locus would reduce the dynamics to a one-locus model. 

\begin{figure}[t!] 
   \centering
   \includegraphics[width= \textwidth]{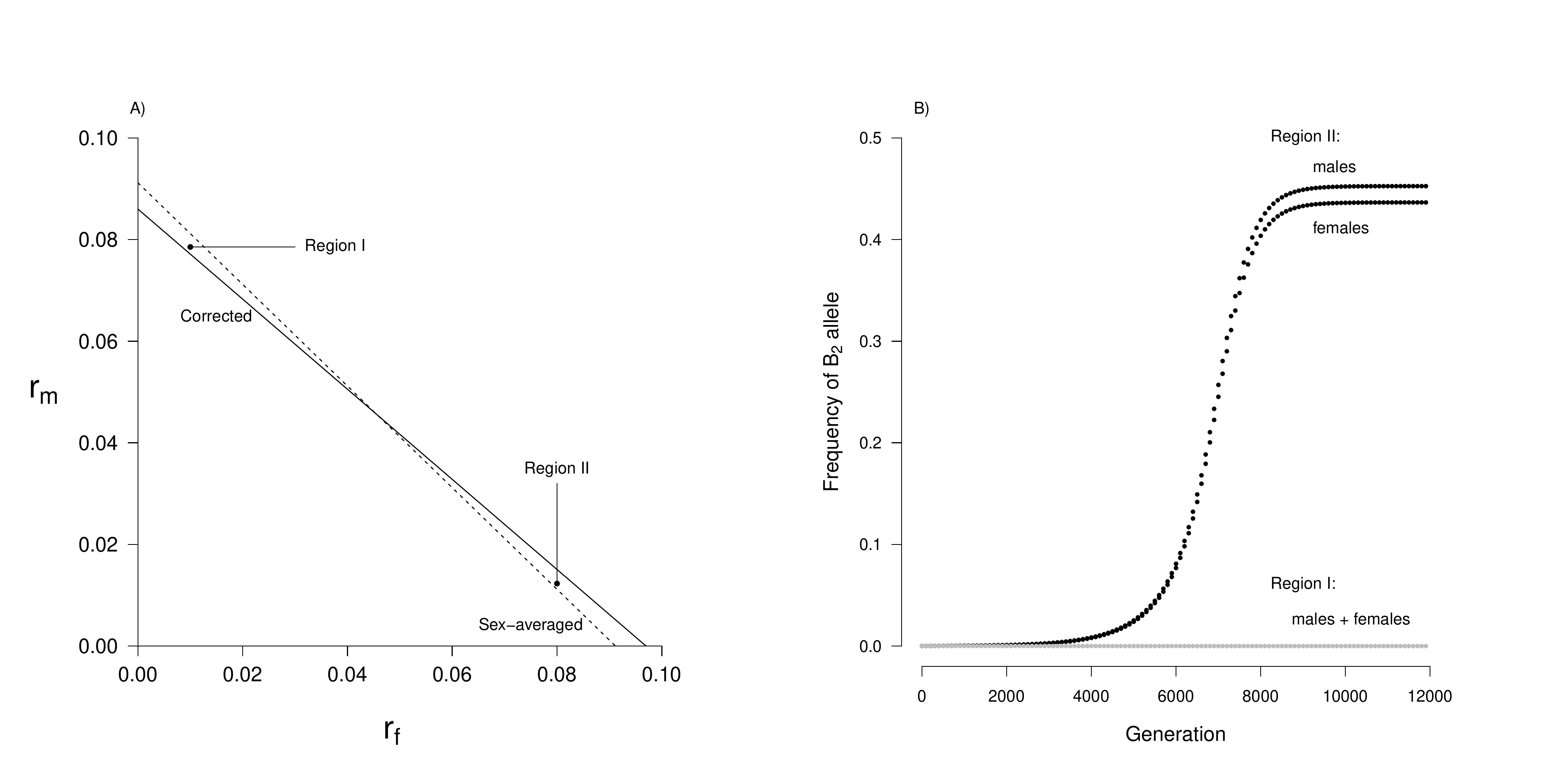} 
   \caption{Sex-specific recombination rates.  See text for parameter selection. a) The dashed line represents the sex-averaged recombination dependent term; it connects the same recombination value on the $r_f$ and $r_m$ axes.  The solid line represents $\lambda_{CC}$ which provides the correction to the sex-averaged estimate. Because the individual recombination rates matter, the solid line does not connect the same values on both axes.  Both lines represent the critical eigenvalue of one. Invasion occurs to the right of the each line; invasion does not occur to the left. In order for recombination to facilitate the invasion of the second male-benefit, female-detriment allele, the range of male recombination rates has to be smaller than the range of female recombination rates.  However, within the invasion space, $r_m = r_f$, $r_m > r_f$ and $r_m < r_f$ can all permit invasion.  b) In region II, the recombination rates ($r_m=0.08$, $r_f=0.012$) allow invasion of $B_2$ even though the sex-averaged rate would not predict invasion.  In region I the recombination rates ($r_m=0.01$, $r_f=0.079$) do not allow invasion of $B_2$, even though the sex-averaged rate would predict invasion. }
   \label{fig1ab}
\end{figure}

In Fig. 1a demonstrates a few features of the model. The solid line is the critical invasionary eigenvalue of one; invasion occurs to the left, and not to the right, of this line.  First, we observe that the range of recombination rates required for the new $B_2$ allele to invade due to its epistatic interaction with $A_2$, differs between male and females.  Males have a smaller range of recombination rates conducive to invasion than females. So, overall, $r_m < r_f$. Connallon and Clark (2010) also show this result but present it in terms of the wait times for successful co-invasion of epistatically beneficial alleles. An intuitive explanation for this theoretical result rests in the fact that keeping $A_2$ and $B_2$ together in males requires suppressing recombination to some degree.  By contrast, because $B_2$ does not benefit females with an $A_2$ background, a wider range of recombination values that can potentially break up the detrimental combinations benefits females.  In support of this interpretation, when $B_2$  hurts females ($t_m < 0$ and $t_f >0$), as in the case of sexual antagonism, this range of female recombination rates increases.

A second, and key, observation is that within the overall invasion space (left of the solid line), $r_m = r_f$, $r_m > r_f$, and $r_m < r_f$ all yield valid conditions for the successful invasion of $B_2$. This may explain why we might not expect to see a consistent fine-scale correlation between sex-specific recombination rates and the concentration of male- or female-benefit alleles -- even though, the range of invasionary recombination rates differs between the sexes. So while the male recombination rate should be lower than the female recombination rate overall in regions enriched with genes beneficial to males, in particular cases they may not be. Conversely, the female recombination rate may not necessarily be lower in regions of female-benefit genes. Positive correlations between the concentration of sex-specific beneficial alleles and lower recombination rates may only be apparent in the most extreme circumstances: e.g., when male and female recombination values are near the invasion border and when males do not recombine (i.e., along the  x-axis of  Fig. 1a).  As Connallon and Clark (2010) have shown, the lack of male recombination and the fact that the X-chromosome experiences selection more often in females than in males may explain empirically why the concentration of male-biased genes is lower on the X chromosome of a male non-recombining species such as \emph{Drosophila}. This must be the case if male-bias can be roughly equated with male-benefit, as the current evidence suggests \pcite[proschel, andolfatto2010, innocentiMorrow, wymanCD].

Finally, Fig. 1a demonstrates that the correction (solid line) to the sex-averaged (dotted line) recombination rate makes the greatest impact near the border of invasion. This borderline region is where the ${\cal O}(u)$ effects and
the ${\cal O}(u^2)$ effects compete with each other to determine invasion. Both lines indicate the slope of the critical invasionary eigenvalue. However, in region II, the sex-averaged rate will incorrectly predict the lack of invasion for this given set of parameters; conversely, in region I, the sex-averaged rate will incorrectly predict successful invasion. The numerical results support this mismatch between the two critical eigenvalues (Fig. 1b). Reliably predicting the persistence or disappearance of novel sexually antagonistic alleles based upon the sex-averaged recombination rate will require that the rate is well within or beyond this borderline.

Many factors lead to variation in the rate of recombination among individuals -- such as age, genetic background, and environmental stress (see \citealt{aucoin} for review).  Sources of variation relevant to this study include re-mating rate \pcite[priest2007] and male-genotype \pcite[stevison]. Sexually antagonistic allele combinations may invade more easily in populations in which males can lower their own recombination rates, or can induce lower average recombination rates in females, relative to the critical invasionary values.  Conversely, male-benefit, female-detriment alleles may invade less easily in populations where males do not decrease (or rather, increase) their own recombination rate or their partner's.  

\subsection{Individual variation in sex-specific recombination rates}

The analyses above support that the sex-averaged recombination rate is not the only quantity of interest.  The sex-specific rates can matter through sexually antagonistic effects on fitness from epistasis. However, because recombination itself is a phenotype, it is also important to consider its variation among individuals \pcite[coopEtAl2008, baudatEtAl2010, fledelAlonEtAl2011] as well as between the sexes. In fact, variation in recombination rate is known to affect variation in fitness; mothers with higher average recombination rate have more children \pcite[kongEtAl2004, fledelAlonEtAl2011].  

To capture these biological realities, our model can be modified so that each sex by genotype combination has a distinct recombination rate, e.g., an adult $A_1B_1A_1B_1$ male has a recombination rate, $r_{11}^m$ (adopting the notation in Table 1). Thus, $A$ and $B$ affect both fitness and recombination rate. In the new eigenvalues of the modified model, recombination only affects terms that carry the double heterozygote genotype (22 = $A_1A_2B_1B_2$; see Table \ref{tab1}):
\ba
\lambda &\supset &  f_{22C} (1-p) (1-r^f_{22C}) + m_{22C} (1-p) (1-r^m_{22C}) \label{heterochiasmyeig} \\
\mbox{or} &&\nonumber \\
\lambda &\supset &  f_{22R} \, p  \, (1-r^f_{22R}) + m_{22R} \, p \, (1-r^m_{22R}). \label{heterochiasmyeig2} 
\ea
This result comports nicely with intuition. Recombination in the double heterozygotes matters because recombination destroys $A_2B_2$ (from the 22C genotype) or creates $A_2B_2$ (from the 22R genotype). Keeping track of these processes will determine the size of the respective eigenvalues. The other genotypes with the $B_2$ allele are either extremely rare (i.e., $B_2B_2$) or have no effective recombination (i.e., recombination occurs but does not change the haplotypes).  

More generally, equations \ref{heterochiasmyeig} and \ref{heterochiasmyeig2} have the form:
\ba
\lambda \supset \mbox{female fitness} + \mbox{male fitness} - (\mbox{female fitness} \times r_f) - (\mbox{male fitness} \times r_m)
\ea
If fitness is always a positive number, and the new allele is male-benefit and female-detriment,  the fourth term is the largest negative term. As a result, invasion becomes more likely (i.e., $\lambda>1$) as $r_m$ becomes smaller. In particular, $r^m_{22C}$ or $r^m_{22R}$ must be smaller, and not any other genotype-specific male recombination rate (e.g., $r^m_{11}$,  $r^m_{12}$,  $r^m_{21}$, etc.). Furthermore, in order for a sexually antagonistic allele to invade, the recombination rate must be low in the sex benefitting from the new allele, and in particular, recombination must be low in the double heterozygote genotype, which carries one copy of the new allele. Interestingly, it is also true that $r^f_{22C}$ or $r^f_{22R}$, and not any other genotype-specific female recombination rate, should be smaller for a male-benefit allele to invade.  

\subsection{Sexually dimorphic allele frequencies}

In the previous sections, we allowed for a polymorphic equilibrium at the $A$ locus, but made the assumption that allele frequencies in males and females were equivalent, $p_{female} = p_{male}$. While assuming equal allele frequencies in the sexes simplifies the math and is a fair approximation, as a precise statement it may have limited biological relevance. After all, when an allele is sexually antagonistic, allele frequencies should be slightly higher in the helped sex and slightly lower in the harmed sex. Although  male and female offspring are equally likely to inherit their alleles from their fathers and mothers,  sexually antagonistic selection and fertility selection can alter allele frequencies prior to mating and fertilization \egcite[pattenEtAl2010, ubedaEtAl2011]. 

To assess and estimate the importance of this mismatch between male and female allele frequencies, we used the recursion equations and found polymorphic equilibria for $p_{male} \neq p_{female}$. The equilibrium allele frequencies can be re-written in terms of two numbers $p_f$ and $d_m$: 
\ba
x_1 = 1 - p_f & \hspace{1 in} y_1 = 1 - p_f - d_m \nonumber \\
x_2 = p_f &  \hspace{1 in}  y_2 = p_f + d_m 
\label{assums}
\ea
where $x_1 = 1-p_{f}$ is the proportion of the ova carrying the $A_1$ allele, $x_2$ the proportion carrying $A_2$. $y_1$ is the proportion of sperm
carrying $A_1$, $y_2$ carrying $A_2$. $d_m$ is the difference in allele frequency between males and females, $d_m$ = $p_m$ - $p_f$. Plugging equations \ref{assums} into the recursion
formulas, using $f = 1 - u \, \delta f$, and  expanding out the mean fitnesses $\bar{w}_m$ and $\bar{w}_f$ and keeping only $u^1$ terms, we found the following approximate equilibrium solutions:
\ba
p_f & = &  \frac{\delta f_{11} - \delta f_{21} + \delta m_{11} - \delta m_{21}}{\delta f_{31} - 2 \delta f_{21} + \delta f_{11} + \delta m_{31} - 2 
\delta m_{21} + \delta m_{11}} \\ \nonumber
\\
d_m & = & \nonumber 2 u \;[\delta f_{31} - \delta f_{21} + \delta m_{31} - \delta m_{21}) (\delta f_{21} - \delta f_{11} + \delta m_{21} -  \delta m_{11}) \times \\&&  \delta f_{11} (\delta m_{31} - \delta m_{21}) + \delta f_{31} (\delta m_{21} - \delta m_{11}) +     \delta f_{21} (-\delta m_{31} + \delta m_{11})] \times \nonumber \\
&& [\delta f_{31} - 2 \delta f_{21} + \delta f_{11} +   \delta m_{31} - 2 \delta m_{21} + \delta m_{11}]^{-3}.
\ea
Although the full expression (to higher order in u) is necessary to find accurate equilibria (see {\tt Mathematica} notebook in online supplement of the journal version of this article for details), this first order approximation has a few instructive properties. $p_f$ appears with no factors of $u$, which makes sense as allele frequencies can be large relative to recombination when $A$ is polymorphic. However, $d_m$ appears as a multiple of $u^1$ (four factors of $\delta f$ or $\delta m$ in the numerator and three in the denominator of $d_m$). Thus, the difference between $p_m$ and $p_f$ has the same size effect on the eigenvalue as the sex-specific fitnesses do. The assumption $p_m = p_f$, in addition to its limited biological appeal, is not necessarily consistent.  

We can calculate how much the eigenvalue is changed relative to $\lambda_{CC}$ when $d_m \neq 0$ by determining the value of  $\Delta \lambda$, defined as
 $\Delta \lambda = \lambda - \lambda_{CC}$ (where $\lambda$ is the full eigenvalue). We can re-write $d_m$ as $d_mu$ to make its size obvious in the counting parameter $u$.
$\Delta \lambda$ is a long expression found by solving for the characteristic polynomial; the most important piece is new first order contribution to the eigenvalue from $d_m$:
\ba
\Delta \lambda = \half d_m \; u + \mathcal{O}(u^2).
\ea
Invasion is substantially impacted by including sexually dimorphic allele frequencies, as represented by $d_m$. Moving onto the next smaller set of terms, we find additional $\mathcal{O}(u^2)$ contributions to the eigenvalue 
that affect $d_m$. Since our main focus is on sexually dimorphic recombination rates, we show only terms with $r_m$ or $r_f$:
\ba
\Delta \lambda  \supset \frac{1}{4}d_mu^2  \(- p_f \,  r_f- 2  r_m+ 
   p_f \, r_m\) + \mathcal{O}(u^3).
\ea
Hence, whenever the sex-specific recombination rates are important (i.e., borderline of Fig. 1) we should also keep track of the new parts of the eigenvalue that are proportional to $d_m$. 

To interpret $d_m$, we can again plug in our fitness model (see Appendix A):
\ba
&p_f&= -\frac{h_f s_f+ h_m s_m}{s_f-2 h_f s_f+s_m-2 h_m s_m}\\
&&\nonumber\\
&d_m&= -\frac{2 (h_f-h_m) s_f s_m ((h_f-1) s_f+(h_m-1) s_m) (h_f s_f+h_m s_m)}{(s_f-2 h_f s_f+s_m-2 h_m s_m)^3}.
\ea
The equation for $d_m$ shows that when male and female dominances are equal, $d_m=0$, so that male and female allele frequencies are identical. Thus, sexual dimorphism in dominance can affect the invasion probabilities, in addition to the effect from sexual dimorphism in recombination rates.

\begin{figure}[t!] 
   \centering
   \includegraphics[width=0.75 \textwidth]{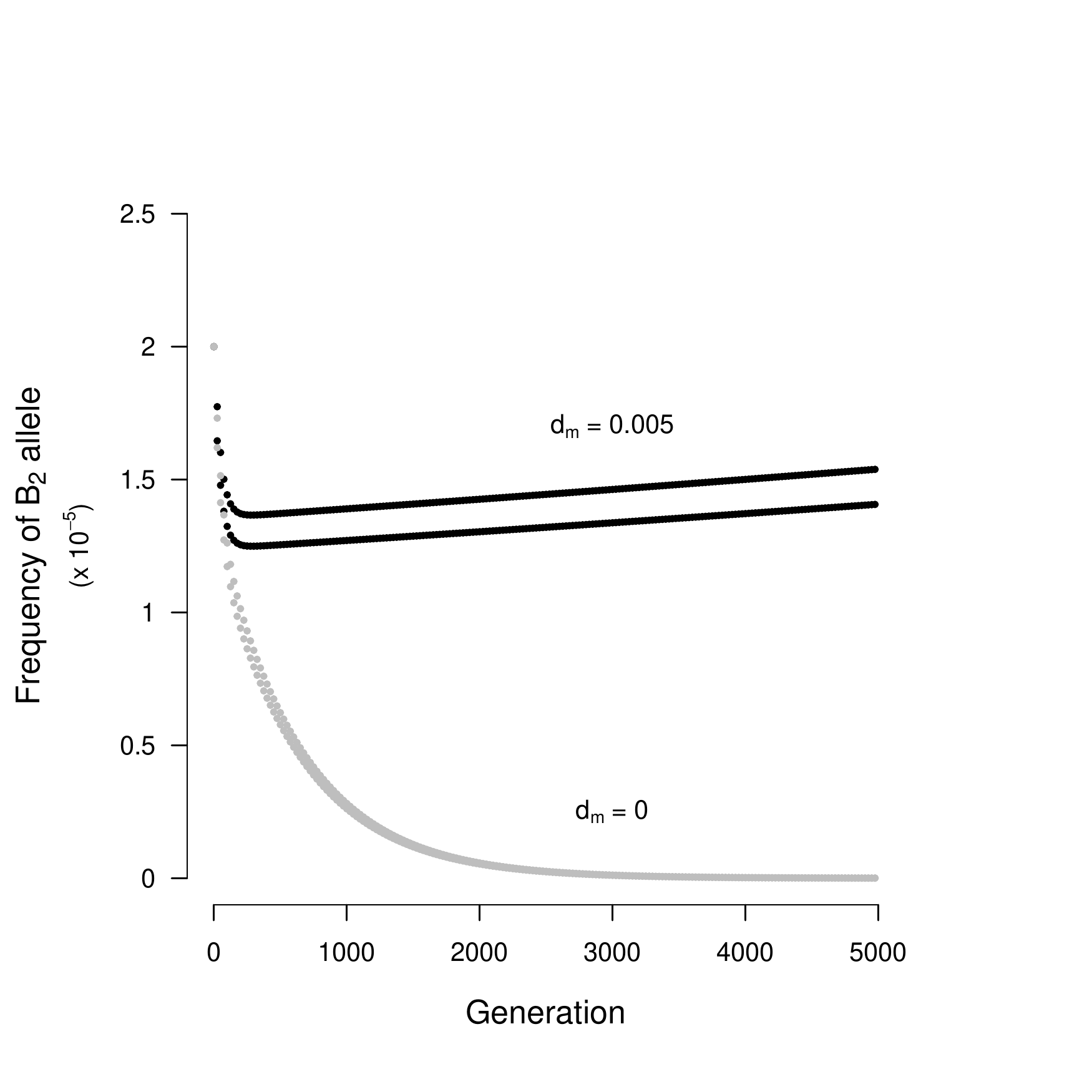} 
   \caption{Differences in allele frequencies of $A_1$ between the sexes. Invasion of the $B_2$ allele (males, upper black points; females, lower black points) can occur in a population which has sexually dimorphic frequencies of $A_1$, with $d_m=0.005$ ($h_m=-0.9$, $h_f=-0.1$, $s_m=0.15$, $s_f=0.05$, $g_m=0.1$, $g_f=0.2$, $t_m=0.05$, $t_f=0.2$, $\epsilon_m=-0.1$, $\epsilon_f=0$). However, invasion of $B_2$ does not occur (both sexes, grey points) when $A_1$ frequency is monomorphic, $d_m\simeq0$ ($h_m=h_f=-0.69$, $s_m=0.1$, $s_f=0.1$, $g_m=0.1$, $g_f=0.2$, $t_m=0.05$, $t_f=0.2$, $\epsilon_m=-0.1$, $\epsilon_f=0$). This difference in invasion outcome exists even though the sex-averaged allele frequencies were initially identical and even though recombination was the same in both populations ($r_m=r_f=0.006$).}
   \label{fig2}
\end{figure}

In  Fig. 2 we numerically confirm that $d_m$ has the potential to impact whether a sexually antagonistic allele $B_2$ can invade in the population. For the same averaged male and female allele frequencies at locus $A$, $B_2$ can eventually invade in one population (black points: $d_m=0.005$) 
but not in the other population (grey points: $d_m=0$).
This occurs despite the fact that both populations have the same recombination parameters ($r_m=r_f=0.006$) and the same sex-averaged $p$. We note that  invasion depends on the size and sign of $d_m$ (larger $d_m$ has a greater impact).  That is, when $d_m>0$ and a new allele is male-benefit,  invasion is  easier. On the other hand, parameter selection that results in $d_m<0$ will hinder the invasion of a male-benefit sexually antagonistic combination (or favor female-benefit alleles). In summary, sex differences in allele frequencies can have definite effects on the possibility of invasion.

\subsection{Limitations of the model}
As with any model, the results are subject to limitations of the assumptions.  Two factors may mitigate the applicability of these results. First, the role of sexually dimorphic recombination was only investigated for weak selection. When the strength of selection increases in either sex, the approximation (equation \ref{ccsimp}) breaks down, and the sex-averaged recombination rate determines invasion outcome over the sex-specific recombination rates.  Second, since populations are finite, it is important to assess the effect of the modified eigenvalues for realistic conditions.  When the eigenvalue is near one (i.e., sex-specific recombination rates make a difference to invasion), we must worry about genetic drift. Drift dominates when $N (\lambda-1) \ll1$, where $N$ is the population size and $\lambda-1$ is the deviation of the eigenvalue for invasion from one. $1/(\lambda-1)$ measures the time to invasion for a novel allele. Hence, in a case when the fitnesses and recombination rates are on the order of 1\%, $(\lambda-1) \sim 10^{-4}$. Drift becomes less important when $N>10^4$, a modest population size.

\section {Conclusions}
Recombination is important for adaptation because it shuffles loci and places favorable alleles together. However, recombination also hinders adaptation by breaking up the very same favorable combinations. The model studied here demonstrates that the advantages and drawbacks of recombination occur alongside each other, albeit in the different sexes.  In fact, sex-specific recombination affects invasion in a manner distinct from simply the sex-averaged recombination rate, especially at the border of invasion. Low recombination permits the helped sex to keep favorable sexually antagonistic combinations together.  Low recombination also permits stable linkage disequilibria to maintain sexually antagonistic variation \pcite[pattenEtAl2010, ubedaEtAl2011]. Both effects occur because low sex-specific recombination increases allele frequencies in the benefitting sex over the harmed sex without the differences washing out over time.

However, within the invasion space, we find that male recombination can be greater than, less than, or equal to female recombination and still allow a male-benefit allele combination to invade. In other words, successful invasion does not always require that the benefitting sex have a lower recombination rate than the harmed sex.  Empirically, the model suggests that low male recombination rates may not always be correlated to a high concentration of male-benefit genes on chromosomes.

When recombination is treated as the phenotype of interest and allowed to vary in a genotype-specific manner, only the recombination rate of the double heterozygotes matters for invasion outcome. The double heterozygotes are the only genotype in which recombination will make a difference to increasing or decreasing the frequency of the new sexually antagonistic allele. This is a sex-specific effect so that a male-benefit, female-detriment allele that simultaneously increases male fitness and decreases male recombination rate invade more easily than those alleles that do only one or the other.

We also find that sexual dimorphism in allele frequencies affects invasion outcomes.  As sexually antagonistic alleles benefit only one sex, allele frequencies may be higher in the helped sex \pcite[pattenEtAl2010, ubedaEtAl2011]. Here, sex differences in allele frequencies at one locus impact invasion of novel alleles at a different locus (Fig. 2).  The magnitude of this effect can be as important as that of sex-specific recombination. These results suggest that maintaining sexually dimorphic allele frequencies at one locus would be a way to both facilitate or hinder the spread of sexually antagonistic alleles at a second locus.

In sum, there is a delicate interplay among sex-specific selection, sex-specific recombination rates, and sex-specific allele frequencies. The impact of sexual dimorphism in recombination rates and allele frequencies became more apparent by doing the calculations than by relying upon intuition alone. Two-locus models of sexual antagonism have a great deal of complexity unavailable to one-locus models that might explain the presence of widespread sexual antagonism in nature. 

\section{Acknowledgments}
We thank L. Rowe, A.F. Agrawal, J.C. Perry, and 2 anonymous referees for comments. M.J.W. was supported by the University of Toronto Connaught Scholarship and Doctoral Completion Award and by NSERC grants to L. Rowe. M.C.W. was supported by  U.S.~Dept.\ of Energy contract DE-FG02-90ER-40560 and by the Kavli Institute for Cosmological Physics at the University of Chicago through grant NSF PHY-1125897 and an endowment from the Kavli Foundation and its founder Fred Kavli.
\pagebreak
\bibliography{recombBib}

\pagebreak
\begin{appendix}

\section{Model of selection and dominance}
\label{supOnline2}
The $m$ and $f$ prefix or subscript indicate male and female, respectively.  The $h$ and $s$ terms describe dominance and selection at the $A$ locus. The $g$ and $t$ terms describe dominance and selection at the $B$ locus.  Epistasis, $\epsilon$, is additive; having two copies of $B_2$ has twice the effect of having one copy. This model assumes \emph{cis-}epistasis since haploid gametes are being modeled; recombination occurs between the $A$ and $B$ locus within a chromosome.\\
\begin{tabbing}
\=$fA_1B_1A_1B_1= 1$ \hspace{2.5in} \= $mA_1B_1A_1B_1= 1$ \\
\>$fA_2B_1A_1B_1= 1-h_fs_f$ \> $mA_2B_1A_1B_1= 1-h_ms_m$ \\
\>$fA_1B_2A_1B_1= 1-g_ft_f$ \> $mA_1B_2A_1B_1=  1-g_mt_m$ \\
\>$fA_2B_2A_1B_1= 1-h_fs_f-g_ft_f-(1/2)\epsilon_f$ \> $mA_2B_2A_1B_1=  1-h_ms_m-g_mt_m-(1/2)\epsilon_m$ \\
\>$fA_2B_1A_2B_1= 1-s_f$ \> $mA_2B_1A_2B_1=  1-s_m$ \\
\>$fA_1B_2A_2B_1= 1-h_fs_f-g_ft_f$ \> $mA_1B_2A_2B_1=  1-h_ms_m-g_mt_m$ \\
\>$fA_2B_2A_2B_1= 1-s_f-g_ft_f-(1/2)\epsilon_f$ \>$mA_2B_2A_2B_1=  1-s_m-g_mt_m-(1/2)\epsilon_m$ \\
\>$fA_1B_2A_1B_2= 1-t_f$ \> $mA_1B_2A_1B_2= 1-t_m$ \\
\>$fA_2B_2A_1B_2= 1-h_fs_f-t_f-(1/2)\epsilon_f$ \> $mA_2B_2A_1B_2= 1-h_ms_m-t_m-(1/2)\epsilon_m$ \\
\>$fA_2B_2A_2B_2= 1-s_f-t_f-\epsilon_f$ \>$mA_2B_2A_2B_2= 1-s_m-t_m-\epsilon_m$\\
\end{tabbing}

\section{Recursion equations and stability analysis}
\label{recursions}

We write the frequency of haplotype $A_i B_j$ in generation $n$ abstractly as $x^{ij} (n)$ (for eggs) and $y^{ij}(n)$ (for sperm). The recursion equations for these frequencies are given by the general equations 
\begin{align}
x^{ij}(n+1)  =& \frac{1}{2 \bar{w}_f} \( x^{ij}(n) \sum_{\ell m} f_{i j \ell m} \; y^{\ell m}(n) + y^{ij}(n) \sum_{\ell m} f_{\ell m i j} \;x^{\ell m}(n)\) - (-1)^{i+j} r_f LD_f\\
y^{ij}(n+1)  =& \frac{1}{2 \bar{w}_m} \( x^{ij}(n) \sum_{\ell m} m_{i j \ell m} \; y^{\ell m}(n) + y^{ij}(n) \sum_{\ell m} m_{\ell m i j} \;x^{\ell m}(n)\) - (-1)^{i+j} r_m LD_m
\end{align}
where $\{i,j,\ell,m\} \in \{1,2\}$, $f_{ij\ell m} (m_{ij \ell m})$ is the fitness of a female (male) with a maternal $A_i B_j$ and paternal $A_\ell B_m$ genotype, $r_f (r_m)$ is the 
recombination rate of females (males), $\bar{w}_f (\bar{w}_m)$ is the mean female (male) fitness, and $LD_f (LD_m)$ is the 
female (male) linkage disequilibrium rate. 
Mean fitness is defined as 
\begin{align}
\bar{w}_f & = \sum_{ijk\ell=1}^2 x^{ij} y^{\ell m} f_{ij\ell m} \\
\bar{w}_m & = \sum_{ijk\ell=1}^2 x^{ij} y^{\ell m} m_{ij\ell m} 
\end{align}
We define
the sex specific linkage disequilibrium factors by
\begin{align}
LD_f &=  \frac{1}{2 \bar{w}_f} \sum_{ij\ell m}^{i\neq\ell, k\neq m} (-1)^{i+j}x^{ij} y^{\ell m} f_{ij\ell m}\\ 
LD_m &=  \frac{1}{2 \bar{w}_m} \sum_{ij\ell m}^{i\neq\ell, k\neq m} (-1)^{i+j}x^{ij} y^{\ell m} m_{ij\ell m} 
\end{align}

If we put the haplotype frequencies in a vector:
\be
X(n)=\{x_{11}(n),y_{11}(n),x_{21}(n),y_{21}(n), x_{12}(n), y_{12}(n), x_{22}(n), y_{22}(n)\},
\ee
then the ${i,j}$ entry of the Jacobian stability matrix $J$ associated with these recursion equations is given by
\be
J_{ij} = \frac{\partial X_i(n+1)}{\partial X_j (n)}.
\ee

The eigenvalues associated with the lower right $4 \times 4$ sub-matrix of $\bf J$ will be the ones relevant to the invasion of the $B_2$ allele, the one relevant to our study. These are reproduced in the main text. We derive them by first finding the characteristic in full generality, and then solve it order by order in an expansion in a small parameter. In our calculations, we do not need to assume weak selection, but always assume weak recombination and a small intersexual fitness difference. That is, we assume all recombination rates are $\ll 1$, and  that the difference between the male and female equilibrium allele frequencies is also small -- i.e., $|p_m-p_f| \ll1$. In the main text of the paper, we have also assumed weak selection when we write most of our equations. These calculations, including equations without the assumption of weak selection, are all reproduced in the associated {\tt Mathematica} notebook
 (see journal version's 
online supplementary materials).

\end{appendix}

\pagebreak
\begin{table}[hp!]
\begin{center}
\caption{Shorthand for genotype fitnesses.}
\begin{tabular}{lll} 
\\
\hline
Diploid genotype & Female fitness & Male fitness \\
\hline
$A_1B_1$ $A_1B_1$ & $f_{11}$ & $m_{11}$\\
$A_1B_1$ $A_2B_1$ & $f_{21}$ & $m_{21}$\\
$A_1B_1$ $A_1B_2$ & $f_{12}$ & $m_{12}$\\
$A_1B_1$ $A_2B_2$ & $f_{22C}$ & $m_{22C}$\\
$A_1B_2$ $A_2B_1$ & $f_{22R}$ & $m_{22R}$\\
$A_2B_1$ $A_2B_1$ & $f_{31}$ & $m_{31}$\\
$A_1B_2$ $A_1B_2$ & $f_{13}$ & $m_{13}$\\
$A_2B_1$ $A_2B_2$ & $f_{32}$ & $m_{32}$\\
$A_2B_2$ $A_1B_2$ & $f_{23}$ & $m_{23}$\\
$A_2B_2$ $A_2B_2$ & $f_{33}$ & $m_{33}$\\
\\
\end{tabular}
\label{tab1}
\end{center}
\end{table}

\section{Online Supplementary Material: eigenvalues to higher order in $r_f$ and $r_m$}
\label{supOnline3}

By performing a perturbative, order-by-order solution of the characteristic polynomial of the recursion equations, we were able to find an expression for the eigenvalue for invasion of the $B_2$ allele with the $A$ locus polymorphic to fourth order in $r$, under the simplifying assumption that
$p_m = p_f$. This is expression is given by:
\begin{align}
\lambda & =  \frac{1}{2\,w_f} \( f_{32} p + f_{22C} q (1-r_f) \) + \frac{1}{2\,w_m} \( m_{32} p + m_{22C} q(1-r_m) \)  \label{fourthOrder}
\\
& + \frac{p q (f_{22C} w_m r_f + m_{22C} w_f r_m) (f_{22R} w_m r_f + m_{22R} w_f r_m)}{2 w_m w_f \( m_{32} w_f p -m_{22R} w_f p +m_{22C} w_f q  - m_{12} w_f q + f_{32} w_m p -f_{12} w_m q+ f_{22C} w_m q- m_{22R} w_m p \)} 
\nonumber \\
&+ \frac{p q (f_{22C} w_m r_f + m_{22C} w_f r_m)(f_{22R} w_m r_f + m_{22R} w_f r_m)((f_{22R}p-f_{22C}q)w_m r_f + (m_{22R}p-m_{22C}q) w_f r_m)}{2 w_m w_f  \( m_{32} w_f p -m_{22R} w_f p +m_{22C} w_f q  - m_{12} w_f q + f_{32} w_m p -f_{12} w_m q+ f_{22C} w_m q- m_{22R} w_m p \)^2}  \nonumber \\
&+ \frac{p q (f_{22C} w_m r_f + m_{22C} w_f r_m) (f_{22R} w_m r_f + m_{22R} w_f r_m)\mathcal{A}}{2 w_m w_f \( m_{32} w_f p -m_{22R} w_f p +m_{22C} w_f q  - m_{12} w_f q + f_{32} w_m p -f_{12} w_m q+ f_{22C} w_m q- m_{22R} w_m p \)^3} \nonumber
\end{align}
where we have rewritten part of the fourth numerator as
\begin{align}
\mathcal{A} &=  w_m^2  ( f_{22C}^2 q^2 + 3 f_{22C} f_{22R} pq + f_{22R} p^2) \, r_f^2  \nonumber \\
&  + w_m w_f (f_{22R} p (3 m_{22C}q+2m_{22R}p)+f_{22C} q (2 m_{22C}q + 3 m_{22R} q)) \, r_f r_m \nonumber  \\
& + w_f^2 (m_{22C}^2q^2 + 3 m_{22C}m_{22R} pq + m_{22R}^2 p^2) \, r_m^2 \nonumber
\end{align}
This extremely complex expression is curious because the denominators for the second, third, and fourth order terms are small numbers, which is not what one would expect. If we write $q=(1-p)$, set the mean fitnesses to one for simplicity, and expand the fitnesses as $f=1-\delta f$, we have
\ba
\mbox{denominator}& \propto& \delta m_{22C}- \delta m_{12} + \delta f_{22C} - \delta f_{12} \nonumber \\
&&+ p (\delta m_{32} - \delta m_{22R}-  \delta m_{22C} + \delta m_{12} + \delta f_{32} -  \delta f_{22R} - \delta f_{22C} + \delta f_{12} ). 
\label{deltaSum}
\ea
All of the $\delta f$ and $\delta m$ terms are small numbers, so their sum (equation \ref{deltaSum}) is also small. Thus, even if the numerators of the each term in the fourth order expansion of equation \ref{fourthOrder} are small, each term is still relatively large. This means that the subsequent higher order corrections to the eigenvalue adjust the eigenvalue by significantly more than one would have expected. Normally,  higher order corrections have a smaller impact. This effect will have a greater impact for larger values of $r$ (since this was an expansion in $r$). In other words, when sex-specific recombination rates are near the borderline of invasion ($\lambda_{CC} \simeq 1$), adding higher order corrections to $\lambda_{CC}$ will affect the slope and location of the line demarcating invasion.

%

 \end{document}